\documentclass{aa}  

\newcommand{\fermi}{{\it Fermi}-LAT}
\newcommand{\gray}{$\gamma$-ray}
\newcommand{\grays}{$\gamma$-rays}
\newcommand{\pks}{PKS 0502+049}
\usepackage{amsmath}
\usepackage{graphicx}
\usepackage{longtable}
\usepackage{multirow}
\usepackage{lscape}
\usepackage[colorlinks=true, citecolor=blue, linkcolor=blue, urlcolor=blue]{hyperref}
\usepackage{txfonts}
\begin{document} 
   \title{The Origin of the Multiwavelength Emission of PKS 0502+049}
   \author{N. Sahakyan
          \inst{1,2}
          }
   \institute{ICRANet-Armenia, Marshall Baghramian Avenue 24a, Yerevan 0019, Armenia.
         \and
             ICRANet, P.zza della Repubblica 10, 65122 Pescara, Italy.
             \email{narek@icra.it}
             }

   \date{Received -; accepted -}

 
  \abstract{The origin of the multiwavelength emission from PKS 0502+049 neighboring the first cosmic neutrino source TXS 0506+056 is studied using the data observed by Fermi-LAT and Swift UVOT/XRT. This source was in a flaring state in the considered bands before and after the neutrino observations in 2014-2015, characterized by hard emission spectra in the X-ray and $\gamma$-ray bands, $\simeq1.5-1.8$ and $\leq2.0$, respectively. During the neutrino observations, the $\gamma$-ray spectrum shows a deviation from a simple power-law shape, indicating a spectral cutoff at $E_{\rm c} =8.50\pm2.06$ GeV. The spectral energy distributions of PKS 0502+049 are modeled within a one-zone leptonic scenario assuming that high energy $\gamma$-ray emission is produced either by inverse Compton scattering of synchrotron or dusty torus photons by the electron population that produce the radio-to-optical emission. Alternatively, the observed $\gamma$-rays are modeled considering inelastic interaction of protons, when the jet interacts with a dense gaseous target. During the neutrino observations, the $\gamma$-ray data are best described when the proton energy distribution is $\sim E_{\rm p}^{-2.61}$ and if the protons are effectively accelerated up to $10$ PeV, the expected neutrino rate is $\sim1.1$ events within 110 days. In principle, if the $\gamma$-ray emission with a hard photon index observed during the flaring periods extends up to $\sim$ TeV, the expected rate can be somewhat higher, but such conditions are hardly possible. Within the hadronic interpretation, the $\gamma$-ray data can be reproduced only when the accretion rate of PKS 0502+049 is in the supper-Eddington regime, as opposed to the leptonic scenario. From the point of view of the necessary energetics as well as considering that the required parameters are physically reasonable, when the neutrinos were observed, the broadband emission from PKS 0502+049 is most likely of a leptonic origin.}

   \maketitle
%
\section{Introduction}\label{sec:1}
The recent observations of Very High Energy ($>100$ GeV; VHE) astrophysical neutrinos by IceCube \citep{2013Sci...342E...1I,2013PhRvL.111b1103A} has opened a new window on studying the nonthermal hadronic processes in the Universe. The neutrino events are distributed isotropically on the sky, suggesting they are of an extragalactic origin. Different source candidates and scenarios have been proposed to explain the origin of the observed neutrinos (e.g., see \citet{2016MNRAS.455..838K,2016PhRvL.116g1101M,2016PhRvD..93h3005W} and \citet{2015RPPh...78l6901A} for a review) but none of them has so far been statistically supported by the observational data.\\
The blazar sub class of active galactic nuclei is often considered as the most likely sources of VHE neutrinos. Such a consideration is natural considering the blazars are among the most luminous and energetic sources in the Universe. Blazars have two jets ejected in opposite directions, one of which is pointing towards the Earth and they are usually sub-grouped into flat spectrum radio quasars (FSRQs) and BL Lac objects, depending on the emission line properties \citep{urry}. The small inclination angle and the relativistic motion in the blazars jets substantially increase their apparent luminosity, so that their emission can be detected across the entire electromagnetic spectrum, from radio to High Energy ($>$ 100 MeV; HE) or VHE \gray{} bands. The non-thermal Spectral Energy Distribution (SED) of blazars has two broad non-thermal peaks - one at the IR/optical/UV or X-ray and the other at HE \gray{} bands. The first peak is due to synchrotron emission of energetic electrons, while the second one can be explained by several different mechanisms. For example, in the so called Leptonic scenarios, the HE emission can be explained by inverse Compton scattering of synchrotron or external photons \citep{ghisellini,ghiselini09,sikora}. Generally, these leptonic scenarios are successfully applied to explain the observed properties in different bands, but sometimes fail to reproduce some observed features such as very fast variability almost in all observed bands (e.g., Mrk 501 \citep{2007ApJ...669..862A} or PKS 2155-304 \citep{2007ApJ...664L..71A} etc.).\\
As an alternative, the HE emission can be explained by the interaction of energetic protons when they are effectively accelerated in the blazar jets. The HE component can be due to proton interaction either with a gaseous target (via proton-proton ($pp$) collisions; \citet{dar, beall, bednarek97}) or with a photon field (proton-$\gamma$ ($p\gamma$) when their energy exceeds the threshold of $\Delta$ resonance \citet{1995APh.....3..295M,1989A&A...221..211M,1993A&A...269...67M,mucke1,mucke2}) or due to proton synchrotron emission \citep{mucke1,mucke2}. The photomeson reaction ($p\gamma$) is more extensively used to explain the emission from blazars \citep{2013ApJ...768...54B}, as it is more likely to have a dense radiation target within the jet than a nuclear one (unless it is of an external origin).\\
Both types of blazars, FSRQs and BL Lacs, are usually considered as effective neutrino emitters. For example, \citet{2016NatPh..12..807K} showed that one of the highest neutrino events detected so far ($\sim2$ PeV) possibly correlates with the bright flare of FSRQ PKS B1414-418. On the other hand, different models (e.g., \citet{2014ApJ...793L..18T, 2015MNRAS.451.1502T}) also predict neutrino emission from BL Lac objects: \citet{padovani2} showed spatial correlation between the extreme BL Lacs (emitting HE \grays{} above 50 GeV) and the arrival direction of the observed neutrino events, once more confirming the blazar-neutrino association.\\
Though blazar have been so far considered as the main sources of VHE neutrinos, no significant association between them and neutrino events has been found yet. The most promising candidate so far is  the blazar TXS 0506+056 (with the coordinates of RA= 77.36 and Dec=+5.69) which can be associated with the neutrino event  IceCube-170922A, detected on 22 September 2017 \citep{IceCubeFermi}. TXS 0506+056 is a bright blazar in the MeV/GeV band at the redshift of z= $0.3365 \pm 0.001$ \citep{2018ApJ...854L..32P}. The multiwavelength observation campaign started after the neutrino alert showed that the source was in an active sate almost in all electromagnetic bands, most interestingly, flaring in the HE and VHE \gray{} bands \citep{IceCubeFermi}. Moreover, IceCube has reported an independently observed $3.5 \sigma$ excess of neutrinos from the direction of TXS 0506+056 between September 2014 and March 2015 \citep{IceCube1}, strengthening the association between the neutrino events and TXS 0506+056. Further, dissection in space, time, and energy of the region around the IceCube-170922A showed that in the \gray{} band the emission from the nearby flaring blazar \pks{} dominates at low energies, but TXS 0506+056 dominates the sky above energies of a few GeV \citep{sah}. Also, during the period of the neutrino excess in 2014-2015, the \gray{} emission from TXS 0506+056 hardened with an excess of hard \gray{} radiation at the highest energies observable by Fermi Large Area Telescope (\fermi{}) \citep{sah}. All these make  TXS 0506+056 the most probable source of the observed VHE neutrinos and many different scenarios have been already proposed to explain the observed neutrinos \citep{2018ApJ...863L..10A,2018arXiv180704275G, 2018arXiv180704335C, 2018ApJ...864...84K, 2018ApJ...865..124M, 2018arXiv180705210L,2018arXiv180900601W,2018ApJ...866..109S}.\\
In this paper, considering the interest toward the region of the sky with TXS 0506+056, the origin of the multiwavelength emission from the neighbouring bright source \pks{} (at $z=0.954$ \citep{1997MNRAS.284...85D}) is investigated using the data from Swift UVOT/XRT and \fermi{} observations. This study is motivated by the fact that \pks{} is only $\sim1.2^{\circ}$ far from TXS 0506+056 and in principle if the neutrinos are produced in the jet of \pks{} they can have some contribution into the IceCube observed events. The aims are: {\it i)} investigation of \pks{} emission properties when VHE neutrinos were observed, using the multiwavelength light curves, {\it ii)} testing of various emission scenarios 
modeling SEDs obtained in different periods and {\it iii)} estimation of the \pks{} neutrino emission rate assuming that the observed HE emission is due to interaction of protons. Such a study will be an independent test if, in the case when hadronic processes are responsible for the HE emission from \pks{}, the produced neutrinos can have any contribution into the events observed by IceCube.\\
The paper is structured as follows. The \fermi{} and Swift UVOT/XRT data analyses are described in Sect. \ref{sec:2}, while the spectral analyses are presented in Sec. \ref{sec:3}. In Sect. \ref{sec:4} the modeling of broadband SEDs within leptonic and hadronic scenarios is presented. The results are discussed and summarized in Sect. \ref{sec:6} .

\section{Observations and Data Reduction}\label{sec:2}
\subsection{Fermi LAT}
\begin{figure*}
  \centering
   \includegraphics[width=0.99 \textwidth]{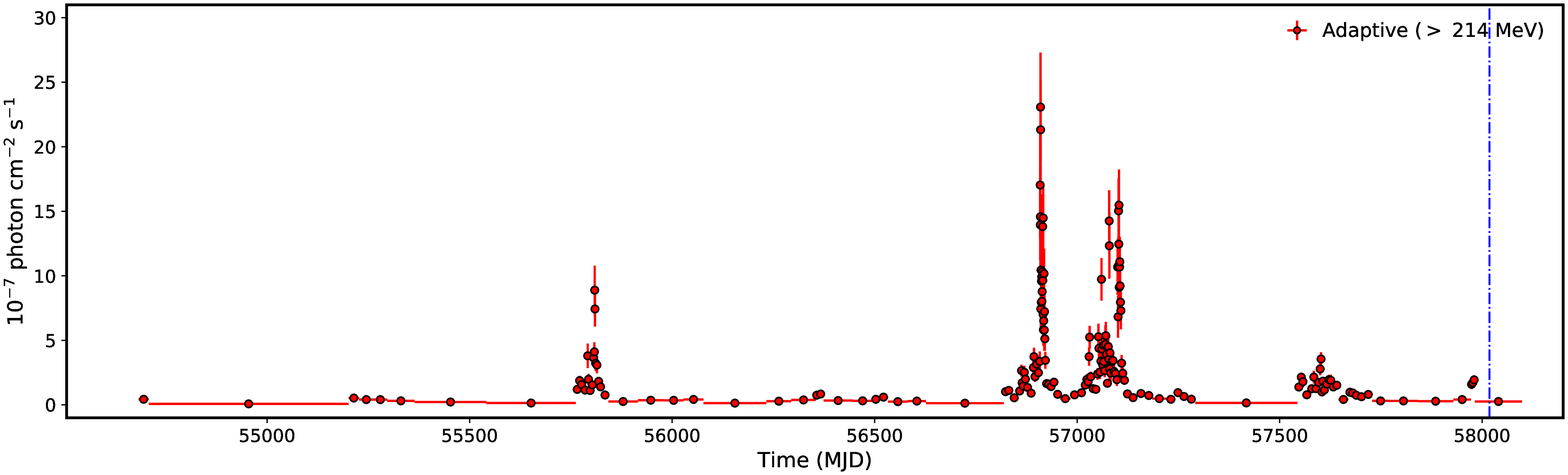}\\
   \includegraphics[width=0.95 \textwidth]{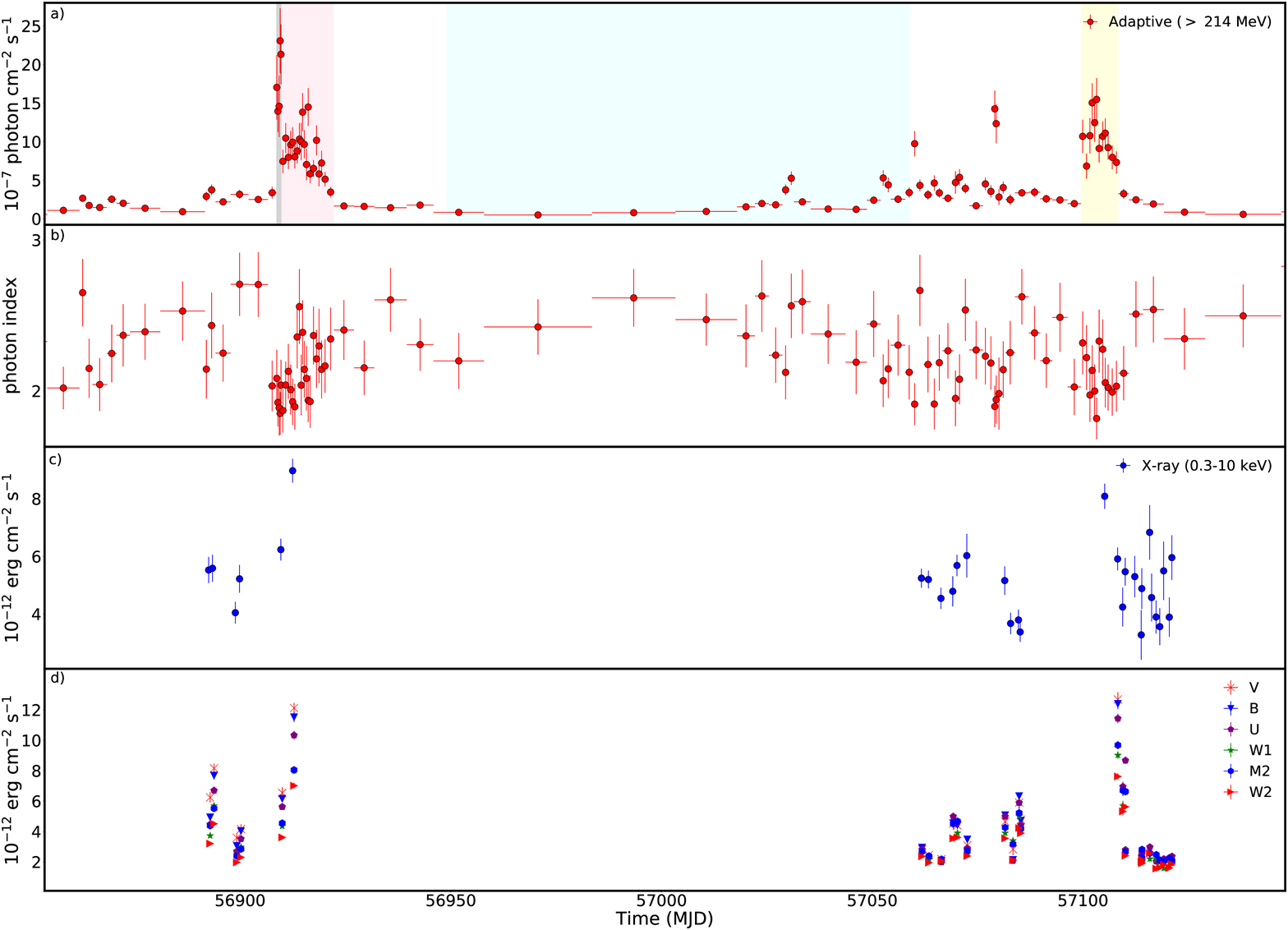}
    \caption{{\it Top Panel:} The \gray{} light curve of \pks{} above 214.0 MeV from August 4, 2008 to January 1st, 2018, with a constant uncertainty of 15\%. {\it Bottom Panels:} (a) the \gray{} light curve and photon index (b), (c) X-ray and (d) optical/UV light curves are shown. The periods P1, P2 and P3 are market with light gray, light red and light yellow colors, respectively, and the period when a $3.5 \sigma$ excess of neutrinos between September 2014 and March 2015 was observed (P0) is in light blue. The blue dot-dashed line shows the period of detection of a HE neutrino event on September 22, 2017.
   }%
    \label{var_mult}
\end{figure*}
\begin{figure}
  \centering
   \includegraphics[width=0.5 \textwidth]{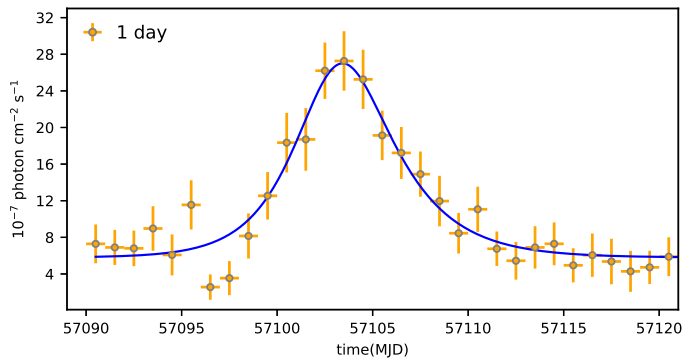}
    \caption{The light curve sub-interval with one-day bins for the flaring period. The blue line shows the flare fit with a double exponential function.
   }%
    \label{flare}
\end{figure}

For the current study the \fermi{} \citep{2009ApJ...697.1071A} data accumulated during more than $9$~years, from 4th August 2008 to 1st January 2018, are used. The 100 MeV - 300 GeV events from a $16.9^{\circ}\times16.9^{\circ}$ square region of interest (ROI) around the \gray{} position of \pks{} (RA,dec)= (76.343, 4.998) were downloaded and analyzed using Fermi Science Tools v10r0p5 with {\it P8R2\_SOURCE\_V6} instrument response function. The events are binned with {\it gtbin} tool into $0.1^{\circ}\times0.1^{\circ}$ pixels and 34 logarithmically equal energy intervals. The standard cuts (e.g., on the maximum zenith angle ($90^{\circ}$) to filter \grays{} from the Earth's limb) are applied with {\it gtselect} and {\it gtmktime} tools. The model file describing ROI was created using \fermi{} 8-year point source list \footnote{https://fermi.gsfc.nasa.gov/ssc/data/access/lat/fl8y/}, including the sources within ROI+$5^\circ$ from the target and Galactic {\it gll\_iem\_v06} and isotropic {\it iso\_P8R2\_SOURCE\_V6\_v06} background models with the normalizations being free parameters. The normalization and spectral indices of the sources within ROI are left as free parameters while for the sources outside the ROI they are fixed to their values obtained during eight years of \fermi{} observations. Then, a binned maximum likelihood analyses is performed with the {\it gtlike} tool. Initially, the spectrum of \pks{} was modeled using a log-parabolic model \citep{2004A&A...422..103M} (as in the \fermi{} catalogs) but for the light curve calculations (for shorter periods) a power-law model was used.\\ 
The  light curve generated by adaptive binning method has been used to investigate the flux variation in time. This novel method allows to identify not only different active states of the source but also find rapid changes in the \gray{} band. The considered period was divided into short (not equal) intervals assuming constant 15\% uncertainty in each bin. The light curve calculated above $E_{0}=214$ MeV optimum energy (for calculation of $E_{0}$ see \citet{lott}) is shown in Fig. \ref{var_mult} (upper panel). The source quiescent state sometimes was followed by rapid and bright flaring periods. The most bright and prolonged \gray{} active period was observed from $\sim$MJD 56900 to MJD 57150, when the highest flux of $F_{>214\:{\rm MeV}}=(2.31\pm0.42)\times 10^{-6}{\rm photon\:cm^{-2}\:s^{-1}}$ was observed on MJD 56909.5 for 4.81 hours. The photon index variation in time is presented in Fig. \ref{var_mult} b) which shows that the flux increase was accompanied by photon index hardening, the hardest one being $1.82\pm0.14$ significantly different than the photon index averaged over nine years ($2.33\pm0.02$). This photon index is unusual for FSRQs which typically have a soft photon index in the MeV/GeV band but for several FSRQs occasionally such hard photon index was observed during rapid flares \citep{2017MNRAS.470.2861S, 2014ApJ...790...45P, 2018ApJ...863..114G}.\\
Then, the light curves during the flares are further analyzed. The flare rise and decay profiles could be constrained only for the bright period around MJD 57100 (see the light curve with one day bins in Fig. \ref{flare}). The flare time profiles are analyzed using the double exponential form function given in \citet{abdoflares} and the fit results are shown in Fig. \ref{flare} with blue line. The rise and decay times of the flare are $t_{\rm r}=2.00\pm 0.35$ days and $t_{\rm d}=2.62\pm 0.39$ days, respectively, with the flare peak at $t_{\rm p}=t_{0}+t_{\rm r}\:t_{\rm d}/(t_{\rm r}+t_{\rm d})ln(t_{\rm d}/t_{\rm r})=$ MJD 57103.43. The constant level present in the flare is $(5.80\pm0.39)\times 10^{-7}{\rm photon\:cm^{-2}\:s^{-1}}$ with the peak flux of $(4.20\pm0.23)\times 10^{-6}{\rm photon\:cm^{-2}\:s^{-1}}$. 
\subsection{Swift XRT/UVOT observations}
The Neil Gehrels Swift observatory (Swift) \citep{2004ApJ...611.1005G} observed \pks{} thirty-five times during the considered period. All the Swift observations were analyzed using the latest version of Swift data reduction software. The data were reprocessed with the standard filtering and screening criteria with the source- and background- extraction regions being defined correspondingly as a 20-pixel ($47"$) radius circular region and an annulus with inner and outer radii being 51 ($120"$) and 85 pixels ($200"$), respectively, both centered at the source position. For all observations, the count rate was below 0.5 count/s, implying no evidence of pile-up. Because of the small number of counts, the Cash statistic \citep{1979ApJ...228..939C} on the unbinned data was used. The spectra  were fitted with an absorbed power-law model in the 0.3-10 keV energy band with a neutral hydrogen column density fixed to its Galactic value $8.76\times10^{20}{\rm cm^{-2}}$ using XSPEC v12.9.1a \citep{1996ASPC..101...17A}.\\
The Swift XRT light curve is shown in Fig. \ref{var_mult} c) and the corresponding parameters are given in Table \ref{tabeswift}. Although the number of available observations is not sufficient for detailed temporal analyses, the X-ray flux increase during the bright \gray{} periods can be noticed. The highest X-ray flux of $(8.91\pm0.42)\times10^{-12}\:{\rm erg\:cm^{-2}\:s^{-1}}$ was observed on MJD 56912.78. No significant spectral evolution was observed in the X-ray band, the photon index most of the time being very hard $\sim(1.2-1.6)$ and the softest one being $\Gamma_{\rm X}\simeq1.87\pm0.39$.\\
The Swift UVOT data have also been analyzed. The source counts were extracted from a circular region of a five-arcsec radius centred on the source, while the background counts from a surrounding annulus (source-free region) with the inner and outer radii being $27"$ and $35"$, respectively. Counts were converted to fluxes using {\it uvotsource} tool and zero-points from \citet{breeveld}. The magnitudes were corrected for extinction, using the reddening coefficient E(B-V) from \citet{schlafly} and the ratios of the extinction to reddening $A_{\lambda}/E(B-A)$  for each filter from \citet{fitzpatrick}, then converting to fluxes following \citet{breeveld}. The averaged flux in Swift UVOT bands is given in Table \ref{tabeswift} and shown in Fig. \ref{var_mult} d). During the \gray{} bright periods also the optical/UV flux has increased.
\begin{sidewaystable*}
\caption{Summary of Swift XRT and UVOT observations of \pks{}} \label{tabeswift}
\centering
\begin{tabular}{cccccccccc} 
\hline
\hline             
Sequence No.& Log(Flux)\tablefootmark{a} & $\Gamma$ & C-stat./dof & $V$\tablefootmark{b} &  $B$\tablefootmark{b} & $U$\tablefootmark{b} & $W1$\tablefootmark{b} & $M2$\tablefootmark{b} & $W2$\tablefootmark{b}\\
\hline
00038377001  & $-12.08\pm0.06$ & $1.45\pm0.16$ & 0.77(89) & $0.99\pm0.11$ & $0.83\pm0.08$ & $0.96\pm0.06$ & $0.97\pm0.05$ & $1.09\pm0.06$ & $0.87\pm0.04$ \\
00038377002  & $-11.66\pm0.05$ & $1.45\pm0.15$ & 0.88(109) & $- -$ & $- -$ & 1.43$\pm$0.04 & $- -$ & $- -$ & $- -$ \\
00038377003  & $-11.26\pm0.03$ & $1.44\pm0.09$ & 0.80(192) & 6.24$\pm$0.35 & 4.94$\pm$0.23 & 4.47$\pm$0.16 & $3.73\pm0.14$ & $4.38\pm0.20$ & 3.21$\pm$0.12 \\
00038377004  & $-11.25\pm0.04$ & $1.45\pm0.10$ & 0.95(196) & 8.15$\pm$0.30 & 7.68$\pm$0.28 & 6.70$\pm$0.25 & $5.69\pm0.21$ & $5.52\pm0.20$ & 4.51$\pm$0.17 \\
00038377005  & $-11.39\pm0.04$ & $1.54\pm0.11$ & 1.08(144) & 3.59$\pm$0.30 & 3.06$\pm$0.20 & 2.69$\pm$0.15 & $2.51\pm0.14$ & $2.39\pm0.11$ & 1.97$\pm$0.11 \\
00038377006  & $-11.28\pm0.04$ & $1.38\pm0.11$ & 0.93(181) & 4.16$\pm$0.31 & 4.03$\pm$0.22 & 3.52$\pm$0.16 & $2.99\pm0.17$ & $2.84\pm0.10$ & 2.30$\pm$0.11 \\
00033408001  & $-11.21\pm0.03$ & $1.53\pm0.07$ & 1.02(257) & 6.54$\pm$0.36 & 6.16$\pm$0.23 & 5.63$\pm$0.21 & $4.36\pm0.16$ & $4.55\pm0.13$ & 3.61$\pm$0.13 \\
00033408002  & $-11.05\pm0.02$ & $1.54\pm0.06$ & 1.33(343) & 12.12$\pm$0.33 & 11.52$\pm$0.32 & 10.33$\pm$0.29 & $8.00\pm0.22$ & $8.05\pm0.22$ & 7.01$\pm$0.19 \\
00033408003  & $-11.28\pm0.03$ & $1.46\pm0.07$ & 1.24(264) & 2.83$\pm$0.26 & 2.95$\pm$0.22 & 2.82$\pm$0.18 & $2.67\pm0.15$ & $2.69\pm0.07$ & 2.37$\pm$0.11 \\
00033408004  & $-11.28\pm0.03$ & $1.60\pm0.07$ & 0.97(262) & 2.46$\pm$0.16 & 2.12$\pm$0.10 & 2.35$\pm$0.09 & $2.14\pm0.08$ & $2.39\pm0.11$ & 1.95$\pm$0.07 \\
00033408005  & $-11.34\pm0.03$ & $1.48\pm0.10$ & 1.00(193) & 2.18$\pm$0.20 & 2.08$\pm$0.17 & 1.99$\pm$0.13 & $2.07\pm0.11$ & $2.18\pm0.10$ & 2.06$\pm$0.09 \\
00033408006  & $-11.32\pm0.04$ & $1.65\pm0.13$ & 0.87(114) & 4.52$\pm$0.42 & 4.54$\pm$0.29 & 4.99$\pm$0.18 & $4.52\pm0.17$ & $4.51\pm0.29$ & 3.55$\pm$0.20 \\
00033408007  & $-11.25\pm0.03$ & $1.65\pm0.08$ & 1.16(227) & 4.40$\pm$0.24 & 4.42$\pm$0.16 & 4.59$\pm$0.17 & $3.90\pm0.14$ & $4.68\pm0.22$ & 3.61$\pm$0.13 \\
00033408008  & $-11.22\pm0.05$ & $1.22\pm0.14$ & 1.02(115) & 3.10$\pm$0.29 & 3.48$\pm$0.22 & 2.90$\pm$0.19 & $2.48\pm0.16$ & $2.74\pm0.18$ & 2.39$\pm$0.13 \\
00033408009  & $-11.09\pm0.02$ & $1.85\pm0.07$ & 1.36(244) & 30.72$\pm$0.85 & 36.08$\pm$0.66 & 33.9$\pm$0.94 & $26.75\pm0.49$ & $24.76\pm0.68$ & 22.18$\pm$0.41 \\
00033408010  & $-11.23\pm0.03$ & $1.57\pm0.08$ & 1.05(216) & 12.69$\pm0.47$ & 12.40$\pm$0.34 & 11.43$\pm$0.32 & $9.02\pm0.25$ & $9.68\pm0.27$ & 7.62$\pm$0.21 \\
00033408011  & $-11.37\pm0.06$ & $1.57\pm0.19$ & 1.25(64) & $- -$ & $- -$ & 6.95$\pm0.32$ & $5.69\pm0.31$ & $6.70\pm0.37$ & 5.32$\pm$0.25 \\
00033408012  & $-11.26\pm0.04$ & $1.56\pm0.11$ & 1.54(151) & $- -$ & $- -$ & 8.67$\pm$0.24 & $6.60\pm0.24$ & $6.63\pm0.24$ & 5.62$\pm$0.16 \\
00033408013  & $-11.28\pm0.06$ & $1.52\pm0.16$ & 0.75(87) & $- -$ & $- -$ & 2.79$\pm$0.21 & $2.75\pm0.20$ & $2.67\pm0.22$ & 2.41$\pm$0.13 \\
00033408014  & $-11.48\pm0.10$ & $1.87\pm0.39$ & 1.10(26) & $- -$ & $- -$ & 2.39$\pm$0.18 & $2.07\pm0.15$ & $2.30\pm0.19$ & 2.00$\pm$0.13 \\
00033408015  & $-11.31\pm0.06$ & $1.60\pm0.17$ & 1.13(78) & $- -$ & $- -$ & 2.55$\pm$0.19 & $2.01\pm0.17$ & $2.24\pm0.21$ & 1.91$\pm$0.12 \\
00033408016  & $-11.17\pm0.06$ & $1.21\pm0.14$ & 0.96(97) & $- -$ & $- -$ & 2.67$\pm$0.20 & $2.27\pm0.17$ & $2.82\pm0.23$ & 2.16$\pm$0.14 \\
00033408017  & $-11.34\pm0.07$ & $1.60\pm0.21$ & 0.73(54) & $- -$ & $- -$ & 2.98$\pm$0.22 & $2.20\pm0.22$ & $2.57\pm0.24$ & 2.57$\pm$0.21 \\
00033408018  & $-11.41\pm0.06$ & $1.85\pm0.20$ & 1.19(60) & $- -$ & $- -$ & 2.04$\pm$0.17 & $2.18\pm0.18$ & $2.48\pm0.21$ & 1.56$\pm$0.12 \\
00033408019  & $-11.45\pm0.07$ & $1.51\pm0.21$ & 0.94(58) & $- -$ & $- -$ & 2.04$\pm$0.17 & $1.99\pm0.17$ & $2.12\pm0.20$ & 1.62$\pm$0.12 \\
00033408020  & $-11.26\pm0.07$ & $1.17\pm0.20$ & 1.31(69) & $- -$ & $- -$ & 2.20$\pm$0.18 & $1.58\pm0.16$ & $2.06\pm0.19$ & 1.81$\pm$0.13 \\
00033408021  & $-11.41\pm0.07$ & $1.41\pm0.19$ & 0.98(63) & $- -$ & $- -$ & 2.28$\pm$0.17 & $1.96\pm0.16$ & $2.24\pm0.19$ & 1.61$\pm$0.10 \\
00033408022  & $-11.22\pm0.05$ & $1.48\pm0.15$ & 1.21(98) & $- -$ & $- -$ & 2.37$\pm$0.20 & $1.90\pm0.16$ & $2.08\pm0.19$ & 1.86$\pm$0.12 \\
00033662001  & $-11.29\pm0.04$ & $1.40\pm0.11$ & 1.11(156) & 4.65$\pm$0.34 & 5.07$\pm$0.23 & 4.99$\pm$0.23 & $3.90\pm0.18$ & $4.26\pm0.12$ & 3.55$\pm$0.16 \\
00033662002  & $-11.43\pm0.04$ & $1.66\pm0.13$ & 1.10(126) & 2.83$\pm$0.42 & 2.13$\pm$0.22 & 2.08$\pm$0.15 & $3.40\pm0.09$ & $3.18\pm0.26$ & 2.12$\pm$0.16 \\
00033662003  & $-11.47\pm0.04$ & $1.69\pm0.13$ & 1.14(138) & 4.78$\pm0.35$ & 4.71$\pm$0.26 & 4.43$\pm$0.20 & $4.12\pm0.19$ & $4.19\pm0.23$ & 3.89$\pm$0.11 \\
00033662004  & $-11.42\pm0.04$ & $1.51\pm0.11$ & 1.33(145) & 5.91$\pm$0.38 & 6.33$\pm$0.23 & 5.89$\pm$0.22 & $5.00\pm0.14$ & $5.22\pm0.48$ & 4.23$\pm$0.16 \\
\hline
\end{tabular}
\tablefoot{
\tablefoottext{a}{Flux in 0.3--10 keV in unit of erg cm$^{-2}$ s$^{-1}$}; 
\tablefoottext{b}{Averaged flux in Swift UVOT bands in units of erg $10^{-12}$ cm$^{-2}$ s$^{-1}$.}
}
\end{sidewaystable*}
\section{Spectral Analyses}\label{sec:3}
The spectra obtained in the following periods are used for investigation of the origin of the multiwavelength emission from \pks{}:
\begin{itemize}
 \item[] From MJD 56949.0 to 57059.0 (P0) corresponding to the neutrino observation window \citep{IceCube1}. Swift observations around this period, Obsid: 33408003, 33408004, 33408005 and 33408006 were analyzed by merging them in order to increase the exposure and statistics as they have similar X-ray fluxes and photon indices.
\item[] From MJD 56908.60 to MJD 56909.80 (P1) during the largest \gray{} flaring period with available quasi-simultaneous Swift observation (Obsid: 33408001).

\item[] From MJD 56909.80 to MJD 56922.23 (P2) when the highest X-ray flux was observed (Obsid: 33408002) with a moderate brightening in the \gray{} band.

\item[] From MJD 57099.53 to MJD 57108.42 (P3), corresponding to another bright \gray{} flaring state coinciding with the Swift observation of Obsid: 33408009.

\end{itemize}
%
%
%
%
%
%
%
These periods are marked with light gray, light red, light blue and light yellow colors in Fig. \ref{var_mult} a). The \gray{} spectra were obtained applying an unbinned likelihood analyses method using a power-law model spectrum with the normalization and index considered as free parameters. After obtaining the best-fit values we fix them for the SED calculations.\\
The results are shown in Fig. \ref{sed} and the corresponding parameters in Table \ref{tab:results}. The \gray{} spectrum contemporaneous with the IceCube observational window ($\sim110$ days; gray) follows the same tendency as that averaged over nine years (light gray) while the \gray{} spectra in the active periods (blue, red and magenta) are significantly different. There is an evident curvature in the \gray{} spectra obtained during P0 (see also \citet{2018arXiv180705057L}) so an alternative fit with power law with an exponential cut- off model in the form of $dN/dE\sim E_{\gamma}^{-\alpha}\:\times Exp(-E_{\gamma}/E_{cut})$ and a log-parabolic function in the form of $dN/dE\sim (E_{\gamma}/E_{\rm br})^{-(\alpha+\beta log(E_{\gamma}/E_{\rm br}))}$ were applied to check if the curvature in the spectrum is statistically significant. The models are compared using a log likelihood ratio test: the significance is the square root of twice the difference in the log likelihoods. The first model with $\alpha=2.07\pm0.04$ and $E_{cut}=8.50\pm2.06$ GeV is preferred over the power-law model with a significance of $7.36\sigma$. Also, the second model with $\alpha=2.23\pm0.02$ and $\beta=0.11\pm0.01$ is preferred with a significance of $6.82\sigma$. The curvature in the blazar emission spectra can be due to different reasons. For example, log-parabolic spectra can be formed when the leptons in the jet undergo stochastic acceleration; power law with an exponential cut-off spectrum is expected when the energy distribution of the emitting electrons has a sharp energy upper cut-off because of the efficiency of the acceleration mechanisms. These results show that the \gray{} emission from \pks{} and consequently the spectra of particles responsible for the emission were characterized by a cut-off at tens of GeV when the neutrino events were detected by IceCube. Interestingly, during the flares before (P1 and P2) and after (P3) this period the \gray{} spectra extend up to tens of GeV with a significantly harder photon index, for example $\Gamma=1.88\pm0.06$ during P1 and $\Gamma\simeq2.0$ during P2 and P3, implying that VHE photons are dominating. This substantial hardening might be caused by injection of new (fresh) particles and/or a change in the location of the emission region where the acceleration is more efficient or the cooling is rather slow allowing the particles to reach higher energies. For generating X-ray spectra, again Cash statistic on Swift unbinned data was applied. Then, in order to increase the significance of individual points in the SEDs calculations, a denser rebinning was applied, restricting the energy range to $>$ 0.5 keV. The results of the fit are given in Table \ref{tabeswift} (similar parameters for the merged observations are: $\Gamma_{\rm X}=1.56\pm0.04$,  $F_{\rm, X}(0.3-10\:{\rm keV})=(4.85\pm0.15)\times10^{-12}$ erg cm$^{-2}$s$^{-1}$) and the corresponding spectra are shown in Fig.  \ref{sed}. During the bright \gray{} periods both the optical/UV and X-ray fluxes increased: the observed shape of UVOT data suggest that it corresponds to the HE tail of the synchrotron component while the hard X-ray spectra are due to the second emission component (in the case of leptonic interpretation).
\begin{table}[t!]
\small
 \begin{center}
 \caption{Parameters of \gray{} spectral analysis}\label{tab:results}
 \begin{tabular}{cccc}
 \hline
  Period  & Flux\tablefootmark{a} & Photon Index & $\sigma$ \\
  \hline
  9 years \tablefootmark{b} & $1.19\pm0.04$ & $2.33\pm0.02$ & 105.3\\
  MJD 56949.00-57059.00\tablefootmark{b}  & $4.17\pm0.16$ & $2.23\pm0.02$ & 65.9\\
  MJD 56908.60-56909.80  & $32.88\pm3.37$ & $1.88\pm0.06$ & 28.9 \\
  MJD 56909.80-56922.23  & $17.42\pm0.85$ & $2.08\pm0.03$ & 56.8 \\
  MJD 57099.53-57108.42  & $20.82\pm1.16$ & $2.01\pm0.04$ & 48.0 \\
  \hline
\end{tabular}
\tablefoot{
\tablefoottext{a}{Integrated \gray{} flux in the $0.1-300$ GeV energy range in units of $10^{-7}\:{\rm photon\:cm^{-2}\:s^{-1}}$}; 
\tablefoottext{b}{Estimated from log-parabola model with $\beta=0.08\pm0.01$ and $\beta=0.11\pm0.02$, respectively.}
}
\normalsize
\end{center}
\end{table}
\section{Modeling of Broadband spectra}\label{sec:4}
As it has been already noted, there are two conceptually different mechanisms which can be responsible for the HE component in blazar emission spectra. The theoretical models are generally divided into leptonic and hadronic ones depending on whether the electrons or protons are responsible for the emission. Here, the multiwavelength emission of \pks{} is discussed within both leptonic and hadronic emission scenarios.
\subsection{Hadronic \grays{} and neutrinos}
In the hadronic or lepto-hadronic blazar jets emission scenarios, the relativistic jet material is composed of protons ($p$) and electrons ($e$) that start to emit when accelerated to ultra-high energies. The low-energy component is dominated by direct synchrotron emission of electrons while the HE component is completely or partially formed due to the radiative output of energetic protons. The blazar jets are ideal laboratories where the protons are sometimes accelerated to above $10^{18}$ eV \citep{1989A&A...221..211M} and their energy is converted into electromagnetic power either due to interaction with gaseous \citep{bednarek97, barkov, araudo13, bednarek15,cita} or photon targets \citet{1995APh.....3..295M,1989A&A...221..211M,1993A&A...269...67M, mucke1, mucke2} and/or via synchrotron emission \cite{mucke1, mucke2}. These channels might, in fact, operate simultaneously in a competing way and contribute to the total energy loss of protons.\\
One of the scenarios most widely applied to explain the HE emission component assumes that the protons interact with the photon field of an internal (e.g., synchrotron photons) or external (e.g., disc photon reflected from Broad Line Region (BLR) or from dusty torus) origin. Then, \grays{}, neutrinos and electron-positron pairs ($e^{-},e^{+}$) are produced from the decay of neutral and charged pions. The \grays{} and $e^{-},e^{+}$ pairs will interact and initiate an electromagnetic cascade that will reduce the energy of the electromagnetic component down to energies at which the source becomes transparent to the $\gamma\gamma$ pair production. In this case, the spectra of the produced neutrinos can be well constrained when the data above 100 GeV are present which are missing for \pks{}. Roughly, assuming that in the $p\gamma$ interactions comparable energy is released into the electromagnetic component (from X- to \grays{}) and  neutrinos, $\phi_{\gamma}\simeq 4 \: \phi_{\nu}$ (e.g., \citet{2016ApJ...831...12H, 2005APh....23..537H}), some constraints on the expected neutrino flux can be imposed. For the X- to \gray{} emission spectrum in the form of $d N_{\gamma}/d E_{\rm \gamma}=N_{0,\gamma}(E_{\gamma}/100\:{\rm eV})^{-\Gamma_{\gamma}}\:Exp(-E_{\gamma}/E_{cut})$ (the power-law spectrum gives poor modeling), the energy flux carried by the electromagnetic component is $\phi_{\gamma}\simeq9.81\times10^{-11}\:{\rm erg\:cm^{-2}\:s^{-1}}$ estimated by fitting the observed data. In this case, the differential spectrum of the accompanying neutrinos can be estimated from $dN_{\nu}/dE_{\nu}=(2-\Gamma_\nu)/(E_{\nu,2}^{2-\Gamma_{\nu}}-E_{\nu,1}^{2-\Gamma_{\nu}})\phi_{\gamma}\:E_{\nu}^{-\Gamma_\nu}$ which predicts a flux of $\simeq6.55\times10^{-16}\:{\rm TeV^{-1}\:cm^{-2}\:s^{-1}}$ at 100 TeV assuming a spectral index of $2.1\pm0.2$, adopting $E_{\nu, min}=1$ TeV and $E_{\nu, max}=10$ PeV. Even if this is a very strict upper limit (the exact estimations require simulations of the proton acceleration and emission processes as well as detailed tracking of cascade propagation) it is already lower than the IceCube measured flux.\\ 
The next scenario for neutrino emission from blazar jets assumes that a dense and compact target (e.g., cloud(s) from BLR \citep{dar,beall,araudo10} or a star/star envelope \citep{bednarek97, barkov, araudo13, bednarek15,cita}) crosses the jet and the accelerated protons penetrating into it interact with the target protons. Depending on the number of jet-crossing targets, the emission can appear as steady (e.g., several clouds can interact with the jet simultaneously) or flare-like. Proton-proton ($pp$) interactions produce neutral ($\pi^0$) and charged pions ($\pi^{\pm}$) which then decay into \grays{} ($\pi^0\rightarrow \gamma \gamma$) and neutrinos ($\pi^+ \rightarrow \mu^+ + \nu_\mu \rightarrow e^+ + \nu_e + \nu_\mu +{\bar \nu}_\mu$).  Unlike the case of $p\gamma$ interaction scenario, a radiation in the MeV/GeV bands is also produced, so the \gray{} data can be used to constrain the proton content in the jet. One of the key points in the jet-target interaction scenario is the acceleration of protons to energies necessary for production of the observed \grays{} and neutrinos; depending on the distance from the base of the jet, where the penetration occurs, the protons can be either accelerated in the jet or in the target when a strong shock is formed, and their energy can go well beyond 10 PeV (a simple relation between the proton acceleration region size $R$ and cooling time scale yields $E_{\rm max}\simeq 3.0\times10^{15}\:(\eta/0.1)\:(B/1{\rm G})\:(R/10^{13}{\rm cm})$ eV \citep{2018ApJ...866..109S}).\\
The jet-target interaction scenario requires several parameters for accurate estimation of the duration, rate and efficiency of interactions. Especially, the parameters describing the target are needed for calculating the related radiative outputs and estimation of the required total  energy of protons. In this case, we do not specify the origin of the dense target and only consider its density indirectly constrained by the observations. Namely, the estimated variability of $t_{\rm v}\simeq2$ days can be used to define the density of the target ($n_{\rm H}$), i.e., comparing it with the characteristic cooling time of $pp$ interactions, $t_{pp}\simeq(K \sigma_{pp} n_{H})^{-1}\simeq10^{15}/n_{H}$, so $n_{\rm H}=5.78\times10^{9}\:{\rm cm^{-3}}$ which is not significantly different from the usually estimated values. As this target density is high, the protons loose a significant fraction of their energy at $pp$ collisions: the interaction is in a radiatively efficient regime, $t_{\rm pp}\leq t_{\rm v}$, so most of the \grays{} are emitted around $t_{\rm v}$ rather than when the target is already accelerated to high velocities.\\
The \gray{} spectra of \pks{} observed in different periods are modeled by expressing the energy distribution of energetic protons as $N_{p}(E_p) \thicksim E_{p}^{-\alpha_{p}} \exp\left(-E_{p}/E_{\rm p, c}\right)$, where the cut-off energy $E_{\rm p, c}$ is initially considered as a free parameter and then fixed to an arbitrary value of $E_{c,p}=10$ PeV; this is selected to ensure the produced neutrinos will have energy above $100$ TeV, but, in principle, a cutoff at much higher energies cannot be excluded. In order to constrain the model parameters more efficiently (the normalization of proton content and their power-law spectral index), i.e., to find the parameters which statistically better explain the observed data,  the Markov Chain Monte Carlo (MCMC) method is employed. This allows to derive the best-fit and uncertainty distributions of the spectral model parameters through MCMC sampling of their likelihood distributions \citep{zabalza}. The neutrino spectra above 100 GeV are calculated following \citet{kelner_06} while at lower energies a delta function approximation is used (for exact formula see \citet{2014ApJ...780...29S}).\\
In the inset of Fig. \ref{sed}, the data observed during P0, P1 and P2 are modeled as \grays{} from the decay of neutral pions ($\pi^0$). During the neutrino observation in 2014-2015, when the power-law index and cut-off in the proton spectrum are considered as free parameters, the data are best described when $\alpha_{\rm p}=2.60\pm0.06$ and $E_{\rm p, c}=3.36\pm2.95$ TeV. The power-law index is mostly defined by the observed \gray{} photon index, whereas the cut-off with a large statistical uncertainty is constrained by the last point in the \gray{} spectrum. When the cutoff is fixed to much larger values, $E_{\rm p, c}=10$ PeV (solid gray line), the data can be reproduced when $\alpha_{\rm p}=2.61\pm0.06$ which predicts also an emission beyond the observed \gray{} data. Due to the steep spectrum of emitting protons, the \gray{} emission is dominated by the decay of $\pi^0$ with a negligible contribution from secondary particles produced by the decay of charged pions. On the other hand, such a steep spectrum also disfavors the possibility of producing a detectable flux of VHE neutrinos. The hardest power-law index when the observed data can be still explained is $\alpha_{\rm p, c}=2.2$ (gray dot-dashed line); however, this will heavily overpredict the \gray{} data above $\sim2$ GeV. Within the applied scenario, the \gray{} spectra observed during the bright P1 and P2 periods can be also modeled (blue and solid lines) when harder indices of $\alpha_{\rm p}=2.14\pm0.10$ and $\alpha_{\rm p}=2.23\pm0.07$ are considered, respectively. Again, the cut-off energy cannot be constrained by the data and, in principle, strong emission of \grays{} and neutrinos up to VHEs can be expected.\\
In this interpretation the total energy of protons (above 1 GeV) in the jet as well as their luminosity can be estimated. Defining the luminosity as $L_{pp}=W_{\rm pp}/t_{\rm pp}$ where $W_{\rm pp}=\int E_p N_{p}(E_p)dE_p$ is the total proton energy integrated from $E_{\rm p, min}$ to $E_{\rm p, max}$ and $t_{\rm pp}=2$ days is the cooling time of protons, the \gray{} data averaged over the IceCube observational window can be modeled when $L_{\rm pp}\simeq1.60\times10^{49}\:{\rm erg\:s^{-1}}$. This luminosity can be as large as $L_{\rm pp}\simeq2.60\times10^{50}\:{\rm erg\:s^{-1}}$ when the \gray{} active periods are considered. These estimations show that if the \grays{} from \pks{} are indeed produced in $pp$ interactions then its jet should be very powerful and efficient in order to transfer a large amount of energy to protons.\\
Constraining the energy distribution of protons and their luminosity, the differential spectrum of the accompanying neutrinos can be calculated straightforwardly. Then, the number of neutrinos detected in a certain exposure of $t_{\rm exp}$ can be estimated from $N_{\nu}\simeq t_{\rm exp} \int A_{\rm eff}(E_{\nu})d N_{\nu}/d E_{\nu} dE_{\nu}$, using the effective area $A_{\rm eff}(E_{\nu})$ from \citep{artsen17}. The neutrino rate ($>200$ GeV) expected within $\sim110$ days can be as large as $\sim1.1$ events when the energy distribution of protons follows $E_{\rm, p}^{-2.61}$ with a cutoff at $10$ PeV. In principle a higher rate ($>20$) is possible when $\alpha_{\rm, p}=2.2$ is considered but in this case the \gray{} data above $1-2$ GeV cannot be explained. This is similar to the case applied in \citet{2018arXiv180804330H} where again the \gray{} emission from \pks{} was interpreted within a jet-target interaction scenario but using a harder proton index. As in this case, the \gray{} data are not well explained when $\alpha_{\rm p}\leq2.0$ which is natural considering the observed steep spectrum in the \gray{} band;  when $pp$ interaction is considered, the produced \grays{} will have nearly the same spectra as those of parent protons, $\alpha_{\gamma}\simeq\alpha_{p}-0.1$. In principle, a hard power-law index of the protons is possible when normalizing it with the sub-GeV \gray{} data, but then a sharp cutoff will be required to describe the observed break at $E_{\rm c, \gamma}=8.50\pm2.06$ GeV. Even at the most unrealistic case when $E_{\rm c,p}=10^4\times E_{\rm c, \gamma}$, the neutrino spectrum, $\sim E_{\nu}^{-\alpha_{\nu}}\exp(-\sqrt{ E_{\nu}/ E_{\nu, c}})$ where $E_{\nu,c}\simeq E_{c,p}/40$ \citep{2007ApJ...656..870K}, will drop above $\sim2.1$ TeV predicting almost no VHE neutrinos. Also, the expected number of neutrinos is somewhat uncertain when the \gray{} active periods are considered, as it strongly depends on the energy cut-off which is unknown. For example, when the cut-off at 10 PeV is considered, the neutrino rate is 14.7 and 22.1 during P1 and P2, respectively, while in the case of $\sim10$ TeV it is as low as $\sim0.75$. This makes any possible claim for neutrino detection during the active periods significantly uncertain.
\subsection{Leptonic HE \grays}
\begin{figure}
   \includegraphics[width=0.49 \textwidth]{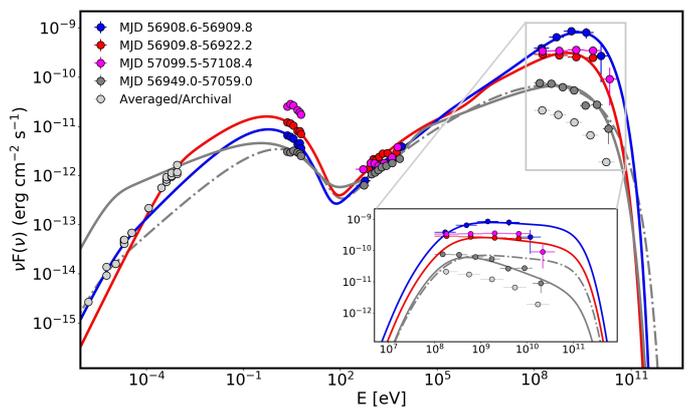}
    \caption{The SEDs of \pks{} during the IceCube observational window (P0; gray) and active states P1 (blue), P2 (red) and P3 (magenta). The averaged \gray{} spectrum during the considered $9$ years and the archival low energy data from ASI science data center are shown in light gray. Gray, blue and red solid lines show the models when inverse Compton scattering of synchrotron (SSC) and torus (EIC) photons are considered, while the gray dot-dashed line is the fitting only with the SSC component. The model fit parameters are given in Table \ref{parres}. The inset shows the \gray{} spectra from $pp$ interactions where the solid lines are the modeling when the cut-off energy in the proton spectrum is fixed to 10 PeV and the gray dot-dashed line is the case when the hard spectrum of protons is considered. The axes are the same as in the main plot. All models have been corrected for $\gamma\gamma$ absorption by the extragalactic background light using the model of \citet{2011MNRAS.410.2556D}.
   }%
    \label{sed}
\end{figure}

In the view of the problems in the hadronic scenarios applied (e.g., the required energetics), the observed broadband emission from \pks{} is discussed also within a leptonic scenario. The multiwavelength spectra for different periods are shown in Fig. \ref{sed} where the archival radio-optical data from ASI science data center and the \gray{} spectra averaged over nine years are shown in light-gray. The spectra in the period when VHE neutrinos were observed (P0) is shown in gray. During the \gray{} active periods the flux increases in all other bands as well, and both components are shifted to higher energies. Here, in the leptonic interpretations, the broadband emission from \pks{} is modeled within the one-zone synchrotron/synchrotron self Compton \citep{maraschi,bloom, ghisellini} plus external inverse Compton \citep{sikora09,ghiselini09,2000ApJ...545..107B} scenarios. \\
In the framework of one-zone leptonic scenarios, the low energy emission (radio through optical) is described by the synchrotron emission of leptons in the magnetic field ($B$), while the HE component (from X-ray to HE \gray{}) is due to the inverse Compton scattering of internal photons, e.g., synchrotron photons (synchrotron self-Compton [SSC]) or external photons (EIC), e.g., emitted from the IR dusty torus. Within this scenario, it is assumed a spherical region (blob) with a comoving radius $R_{\rm b}$ is moving with a bulk Lorentz factor $\Gamma_{\rm b}$ toward the observer and is filled with an isotropic population of electrons and a randomly oriented uniform magnetic field $B$. The energy spectrum of the injected electrons in the jet frame can be expressed as (e.g., \citep{1996ApJ...463..555I})
\begin{equation}
N^{\prime}_{\rm e}(E^{\prime}_{\rm e})= N^{\prime}_{0}\:\left( E^{\prime}_{e}/m_{e}\:c^2\right)^{-\alpha}\:Exp[-E^{\prime}_{\rm e}/E^{\prime}_{\rm cut}]
\label{elect}
\end{equation}
for $E^{\prime}_{\rm min}\leq E^{\prime}_{\rm e}\leq E^{\prime}_{\rm max}$ where $E^{\prime}_{\rm min}$ and $E^{\prime}_{\rm max}$ are the minimum and maximum electron energies, respectively. The emitted radiation will be Doppler-boosted by $\delta$ which equals to the bulk Lorentz factor for the small jet viewing angles. For the Doppler factor, a typical value of $20$ \citep{ghistav} will be adopted which is usually used for the modeling of emission from FSRQs. The radius of the emission region can be  constrained by the variability time scales: the radius can not be larger than $R_{\rm b}\leq c\times t \times \delta/(1+z)\simeq5.31\times10^{16}\:(\delta/20)\:{\rm cm}$.\\
Usually, the Compton dominance (domination of the second emission peak) observed from FSRQs can be explained by inverse Compton scattering of the external photon fields. If the jet dissipation occurs within the BLR whit a radius of $7.6\times10^{17}\:{\rm cm}$ for \pks{} (measured using $R_{\rm BLR}\sim\lambda L_{\lambda}(5100\AA)^{0.7}$ relation \citep{2002ApJ...576...81O}) the dominant external photon fields are disc photons reflected by the BLR clouds. On the other hand, the recent observations in the VHE \gray{} band indicate that the emission region can be also well beyond the BLR where the dominant photon field is IR radiation of the dust tours \citep{2015ApJ...815L..22A, 2015ApJ...815L..23A, 2011ApJ...730L...8A}. These regions appear more favorable for the VHE \gray{} emission (e.g., \citet{2018ApJ...863..114G}). In the current study the torus photons are taken into account assuming the emission from the torus has a blackbody spectrum with a temperature of $T = 10^3$ K and fills a volume that for simplicity is approximated as a spherical shell with a radius of $R_{\rm IR} = 3.54 \times 10^{18} (L_{\rm disc}/10^{45})^{0.5}$ cm \citep{2008ApJ...685..160N}. The corresponding radiation energy density, as measured in the comoving frame would be $u_{\rm torus}=\eta L_{\rm disc} \delta^2/4 \pi R_{\rm torus}^2 c\simeq5.1\times10^{-2}\:(\delta/20)^2\:{\rm erg\:cm^{-3}}$ where $\eta = 0.6$ \citep{2009MNRAS.397..985G}. During the fitting,  the model free parameters (magnetic field and parameters describing the nonthermal electron distribution) and their uncertainties are estimated applying the MCMC method using {\it naima} package \citep{zabalza}.\\
\begin{table}[t!]
\small
 \begin{center}
 \caption{Parameters of \gray{} spectral analysis}\label{parres}
 \begin{tabular}{llll}
 \hline
 \hline
   & P0 & P1& P2 \\
  \hline
  $\alpha$ & $1.82\pm0.02$ & $1.61\pm0.05$ & $1.90\pm0.07$ \\
$E_{\rm min}^{\prime}$[MeV] & $12.97\pm7.17$ & $14.71\pm8.59$ & $65.68\pm23.70$ \\  
$E_{\rm c}^{\prime}$[GeV] & $7.76\pm0.39$ & $2.51\pm0.20$ & $1.99\pm0.14$ \\
$E_{\rm max}^{\prime}$ [TeV] & $0.63\pm0.29$ & $0.58\pm 0.35$ & $0.68\pm0.28$ \\
$B$[mG] & $31.27\pm0.61$ & $102.92\pm8.03$ & $235.93\pm8.17$\\
$L_{\rm B}^{\prime}[{\rm erg\:s^{-1}}]$ & $4.13\times10^{42}$ & $4.48\times10^{43}$ & $2.35\times10^{44}$ \\
$L_{\rm e}^{\prime}[{\rm erg\:s^{-1}}]$ & $1.87\times10^{46}$ & $7.52\times10^{45}$ & $4.26\times10^{45}$\\
\hline
\end{tabular}
\normalsize
\end{center}
\end{table}
The modeling of SEDs observed during, P0, P1 and P2 are shown in Fig. \ref{sed} and the corresponding parameters are given in Table \ref{parres}. In all modeling, the radio data are not considered as they are not simultaneous and the emission in this band can be produced from the low-energy electrons in more extended regions. Initially, the HE component observed during P0 is modeled considering only SSC mechanisms (gray dot-dashed line) as due to compactness of the emitting region the density of synchrotron photons might be dominating. The observed data are relatively well explained when $E^{'}_{\rm min}=12.97\pm7.17$ MeV, $\alpha=1.82\pm0.02$ and $E_{\rm c}^{\prime}=7.76\pm0.39$ GeV. However, as the HE component exceeds that at lower energies, this modeling requires a strongly particle-dominated jet $U_{\rm e}/U_{\rm B}\simeq4.5\times10^{3}$ for $B=31.27\pm0.61$ mG. The required extreme parameters can be softened when the contribution from external photons is considered. For example, the  solid gray line represents the modeling of the data considering inverse Compton scattering of both synchrotron and torus photons. This requires a softer power-law index for the electrons $\alpha=2.42\pm0.28$ and as the energy of torus photons exceeds the averaged energy of synchrotron ones, this modeling requires lower minimum and cutoff energies of $E^{'}_{\rm min}=5.91\pm0.61$ MeV  and $E^{'}_{\rm c}=2.91\pm0.21$ GeV, respectively. In this case, the synchrotron emission of the low energy electrons will exceed the observed radio flux a few times but as the radio data are not contemporaneous, this cannot be a strong argument to disfavor such modeling. Unlike the previous case, the system is close to equipartition $U_{\rm e}/U_{\rm B}\simeq19.6$. Similarly, the spectra observed in bright P1 and P2 are modeled considering the SSC and EIC mechanisms. For both periods, the optical/UV and X-ray data can be explained by synchrotron/SSC emission while the \gray{} data are due to the inverse Compton scattering of external photons from dusty torus. During P1 the power-law index of emitting electrons was $\alpha=1.61\pm0.05$ defined by the hard \gray{} photon index, while it was $\alpha=1.90\pm0.07$ during P2 when a nearly flat spectrum in the \gray{} band was observed. The cut-off energy of $E_{\rm cut}^{\prime}=1.99-2.51$ GeV is measured from the optical/UV data which is not significantly different for the two periods. The magnetic field in P2 ($B=235.93\pm8.17$ mG) is slightly larger than that in P1 ($B=102.92\pm8.03$ mG) in agreement with the observed increase in the optical/UV bands. The total luminosity of the jet defined as $L_{\rm jet}=L_{\rm B}+L_{\rm e}$ where $L_{B}=\pi c R_b^2 \Gamma^2 U_{B}$ and $L_{e}=\pi c R_b^2 \Gamma^2 U_{e}$ \citep{2008MNRAS.385..283C} is in the range of $L_{\rm jet}\simeq(4.50-7.56)\times10^{45}\:{\rm erg \:s^{-1}}$. During P1 the jet is particle-dominated with $U_{\rm e}/U_{\rm B}=167.9$ while for P2 $U_{\rm e}/U_{\rm B}=18.1$.
\section{Discussion and Conclusions}\label{sec:6}
Blazar jets have always been assumed as the most promising sources of VHE neutrino emission. The recent association between the IceCube-170922A neutrino event with the \gray{} bright BL Lac object TXS 0506+056 has opened new perspectives for investigation of the blazar jets physics. For the first time, the emission processes in relativistic jets can be studied using both \grays{} and neutrinos. Though there are various arguments favoring TXS 0506+056 as the main source for the observed VHE neutrinos, additional care must be taken when considering the presence of the nearby powerful \gray{} emitter- \pks{}. In this paper the origin of the multiwavelength emission from FSRQ \pks{} is investigated aiming to verify whether or not the possible neutrino emission from \pks{} accompanying the observed \gray{} flux can have contribution to the IceCube observed events. For this purpose, the \gray{} data from \fermi{} and optical/UV/X-ray data from Swift UVOT/XRT observations of \pks{} in 2008-2018 have been analyzed. In the \gray{} band the source showed several bright periods. The maximum flux of $(4.10\pm0.75)\times 10^{-6}{\rm photon\:cm^{-2}\:s^{-1}}$ integrated above 100 MeV was observed on MJD 56909.5 within 4.81 hours. During the highest flux, the apparent isotropic \gray{} luminosity is $L_{\gamma}\simeq4.72\times10^{49}\:{\rm erg\:s^{-1}}$ (using a distance of $d_{\rm L}\simeq 6269.5 {\rm Mpc}$) which corresponds to $L_{em, \gamma}=L_{\gamma}/2\delta^2\simeq5.90\times10^{46}\:{\rm erg\:s^{-1}}$ (when $\delta=20$) total power emitted in the \gray{} band in the proper frame of the jet. The \gray{} photon index varies as well, being very soft during the low states while significantly hardening in the bright periods, the hardest one being $\Gamma=1.82\pm0.14$. In the X-ray band, the flux is of the order of a few times 10$^{-12}$ erg cm$^{-2}$ s$^{-1}$ but with a hard photon index $\simeq1.2-1.6$, unusual for FSRQs. The X-ray flux variation cannot be tested, as there are only few observations; however, an evidence of flux increasing around the \gray{} flares can be seen. Similar tendency is present also in the optical/UV data obtained by Swift UVOT.\\
The \gray{} spectra when VHE neutrinos were observed as well as during the \gray{} active periods were obtained. The curved \gray{} emission spectrum during MJD 56949-57059 is better explained by a power-law model ($\sim E^{-2.07}$) with a cutoff at $E_{cut}=8.50\pm2.06$ GeV. This implies the presence of a cut-off in the energy distribution of the parent population of particles responsible for the emission, so the HE processes were not dominant/efficient in the jet of \pks{} when the neutrinos were observed by IceCube. In this period, the emission from TXS 0506+056 was not dominating in the lower \gray{} band but there is an indication of a hard emission component in the higher-energy \grays{} \citep{sah}, showing that most likely there  was an efficient contribution from the VHE particles. When the active periods before and after the neutrinos observation window are considered, the \gray{} emission from \pks{} appears with a very hard \gray{} photon index of $\leq2.0$. This shows even if there are certain periods when the jet of \pks{} was in a favorable state for HE and VHE \gray{} emissions, it seems not to be the case when neutrinos were observed.\\
Nearly symmetric flare time profiles with the shortest flux e-folding time being $t_{\rm r}=2.00\pm 0.35$ days are obtained for the flare around MJD 57100. The rise and decay of the flare can be explained by acceleration and cooling of electrons. For example, the cooling of electrons of $E_e=1\:{\rm GeV}$ within $t_{\rm d}=2.62\pm 0.39$ day requires a magnetic field of $B\approx 0.30\;{\rm G}\:(\delta/20)^{-1/2}(t_{\rm dec}/2.62\:d)^{-1/2}\:(E_{e}/1 GeV)^{-1/2}$ ($t_{\rm cooling}=\delta\times t_{\rm d}=6\:\pi \:m_e^2\:c^{3}/\sigma_{\rm T} B^2\:E_{e}$) which is typical for blazars.\\
The multiwavelength emission from \pks{} is interpreted within leptonic and hadronic scenarios. In the hadronic interpretations, the absence of VHE \gray{} data prevents exact estimations of expected neutrino rates when $p\gamma$ scenario is considered and only quantitative limits can be imposed. In the most optimistic case, the neutrino flux predicted at 100 TeV falls below the IceCube estimated one, implying the neutrinos accompanying the observed electromagnetic emission (from X- to \gray{} bands) can not be the source of the observed neutrinos. Next, if the observed \grays{} are due to $pp$ interactions in the dense target crossing the jet, then the energy of protons is mostly released in the GeV band allowing a straightforward measurement of the proton spectra based on the observed \gray{} data. The \gray{} data obtained during the IceCube observational window can be well explained when the energy distribution of protons is $E_{\rm p}^{-2.61}$. Then, if the proton cutoff energy is at $\sim10$ PeV, the maximum possible neutrino detection rate will be $\sim 1.1$ events. A higher neutrino detection rate is possible when a harder power-law index of protons $\alpha_{\rm p}=2.2$ is considered; however, it strongly over-predicts the HE \gray{} data above $1-2$ GeV. Alternatively, a significant neutrino emission is expected during the \gray{} flaring periods when $\alpha_{\rm p}=2.1-2.2$ and only if $E_{\rm p,c}\geq 100$ TeV; for example, in order to have a detection rate of $>4.0$ events, it is required that the hard \gray{} spectra extend at least up to $\simeq E_{\rm c,p}/40=2.5$ TeV - these extreme conditions are hardly possible.\\
In the leptonic interpretations, the broadband spectra of \pks{} are modeled within the one-zone leptonic scenario assuming the emission is produced in the compact region ($R\leq 5.31\times10^{16}\:(\delta/20)\:{\rm cm}$ constrained by the observed variability). When the synchrotron/SSC radiation model is considered, the observed data can be explained only when the electron energy density strongly dominates over that of the magnetic field. Instead, the data can be better explained when the inverse Compton scattering of external photons is taken into account; assuming the emitting region is outside the BLR, SSC radiation from the electron population producing the radio-to-optical emission can describe the observed X-ray data while the emission in the \gray{} band with a large Compton dominance can be explained by the IC scattering of dusty torus photons. This interpretation does not require extreme parameters unlike it does in the case of $pp$ interaction scenario, for example, the multiwavelength SED obtained during the IceCube observations can be explained when the electron power-law index is $\alpha=2.42\pm0.28$ above $E^{'}_{\rm min}=12.97\pm7.17$ MeV and $E^{'}_{\rm c}=2.91\pm0.21$ GeV and the emitting region is not far from equipartition $U_{\rm e}/U_{\rm B}\simeq19.6$. Similar parameters required in the modeling of flaring states are $\alpha=1.6-1.9$ and $E_{\rm c}\simeq2.5$ GeV and the magnetic field of $B=(102.9-235.9)\:{\rm mG}$ with an energy density not significantly different from that of the electrons $U_{\rm e}/U_{\rm B}=(18-168)$. The estimated emitting electron parameters are supported by the currently known acceleration theories and the other parameters are physically reasonable.\\
In the leptonic and hadronic modeling the required energetics of the system is significantly different. For example, the estimated luminosity in the leptonic scenario varies within $L_{\rm jet}\simeq(4.5-18.7)\times10^{45}\:{\rm erg \:s^{-1}}$ comfortably below the Eddington luminosity of \pks{} ($L_{\rm Edd}\simeq9.15\times10^{46}\:{\rm erg \:s^{-1}}$ for the black hole mass of $7.53\times10^{8}M_\odot$ \citep{2002ApJ...576...81O}), while in the hadronic interpretation, the accretion should be at super-Eddington rates as the required luminosity exceeds the Eddington limit by $2-3$ orders of magnitude. Although super-Eddington accretion rate is not rare for blazars, it imposes strong difficulties on the hadronic interpretation.\\
In this paper, we attempt to investigate the origin of multi-wavelength emission from \pks{} during the observation of VHE neutrinos in 2014-2015 and of \gray{} flaring periods as well as investigate whether the neutrino emission from \pks{} can have any contribution to the events observed by IceCube. The spectra observed in all periods {\it can be well reproduced by the leptonic models with physically reasonable parameters} unlike the hadronic models which require a substantially higher jet luminosity. Even in these extreme conditions, based on the \gray{} data the expected neutrino rate can be only $\sim1.1$ events. In this view, considering the required energetics and predicted spectral shapes, the nearby blazar TXS 0506+056 is a more preferred source of VHE neutrinos. The presented discussion and modeling show that the broadband emission from \pks{} is most likely of a leptonic origin, leaving TXS 0506+056 as the first extragalactic source of VHE neutrinos.
\section*{Acknowledgments}
This work was supported by the RA MES State Committee of Science, in the frames of the research project No. 18T-1C335.
\bibliographystyle{aa}
\bibliography{references}{}
\end{document}